\documentclass[a4paper,fleqn]{cas-dc}

\usepackage{braket}
\usepackage{hyperref}
\usepackage[numbers]{natbib}
\def\tsc#1{\csdef{#1}{\textsc{\lowercase{#1}}\xspace}}
\tsc{WGM}
\tsc{QE}
\tsc{EP}
\tsc{PMS}
\tsc{BEC}
\tsc{DE}
\begin{document}
\let\WriteBookmarks\relax
\def\floatpagepagefraction{1}
\def\textpagefraction{.001}
\title [mode = title]{Bohm approach to the Gouy phase shift}                      
\author[1]{H\'ector M. Moya-Cessa}
\author[2,3]{Sergio A. Hojman}
\author[3]{Felipe A. Asenjo}
\author[1]{Francisco Soto-Eguibar}
\cormark[1]
\address[1]{Instituto Nacional de Astrofísica Óptica y Electrónica, Calle Luis Enrique Erro No. 1, Santa María Tonantzintla, Pue., 72840, Mexico}
\address[2]{Departamento de Ciencias, Facultad de Artes Liberales, Universidad Adolfo Ibáñez, Santiago 7491169, Chile.\\
Departamento de Física, Facultad de Ciencias, Universidad de Chile, Santiago 7800003, Chile.\\
Centro de Recursos Educativos Avanzados, CREA, Santiago 7500018, Chile.}
\address[3]{Facultad de Ingeniería y Ciencias,
Universidad Adolfo Ibáñez, Santiago 7491169, Chile.}
\cortext[cor1]{feguibar@inaoep.mx}

\begin{abstract}
By adapting the Madelung-Bohm formalism to paraxial wave propagation we show, by using Ermakov-Lewis techniques, that the Gouy phase is related to the form of the phase chosen in order to produce a Gaussian function as a propagated field. For this, we introduce a quantum mechanical invariant, that it is explicitly time dependent despite the fact that the Hamiltonian is itself time-independent. We finally show that the effective Bohm {\it index of refraction} generates a GRIN medium that produces the focusing needed for the Gouy phase.
\end{abstract}


\maketitle

\section{Introduction}
The Gouy phase \cite{Gouy1,Gouy2,Michigan,Michi2,Siegman} is a phase gradually acquired by a beam around the focal region and it results in an increase in the apparent wavelength near the waist ($z \approx 0$) deriving in the fact that the phase velocity in such region formally exceeds the speed of light. It has been explicitly shown that the Gouy phase shift of any focused beam originates from transverse spatial confinement, that introduces a spread in the transverse momentum and therefore a shift in the expectation value of the axial propagation constant. {\color{black}The simultaneous effect of confinement in space combined with spread in momentum is, of course, reminiscent of Heisenberg's uncertainty principle.} \cite{Michigan}. The Gouy phase has been used to  generate arbitrary cylindrical vector beams \cite{Photonics}, to shape vectorially structured light with custom propagation-evolution properties \cite{Carmelo} and to give a new Bateman-Hillion solution to the Dirac equation for a relativistic Gaussian electron beam \cite{Ducharme}, among other applications. Moreover, its origin has been investigated over the years \cite{Michigan,JOSAA,ACS}.

On the other hand, it is well known that during free propagation light  bends when an initial ($z=0$) Airy field is considered \cite{Sivil,Vo,Efrem,Cerda,Jaure,Torre}, which may be explained by using Madelung-Bohm theory \cite{ChileI,ChileII}. This formalism has been applied to solve  the Schrödinger equation for different systems by taking advantage of their non-vanishing Bohm potentials \cite{Hojman,Hojman2,Hojman3,makowski}. We therefore allow to ask if Madelung-Bohm theory \cite{Madelung,Bohm} could be a way to explain Gouy's phase. We show that indeed it may be explained and to this end we use Ermakov-Lewis techniques in order to use the invariant introduced by Lewis for a time dependent harmonic oscillator \cite{Lewis0,Lewis}, and thus use it for a free particle. 

The manuscript is organized as follows: In the next Section, we show that for a free particle an invariant of the Ermakov-Lewis form may be written which, remarkably, is explicitly time dependent. We then give a solution to the Ermakov equation for this particular case. In Section 3, by taking advantage of the similarity between the Schrödinger equation and the paraxial wave equation, we translate the Madelung-Bohm theory to classical optics. In Section 4, we give an operator solution to the (optical) Bohm equations and propose a Gaussian initial field. By using the Ermakov solution obtained in Section 1, we calculate the Gouy phase. Section 5 is left for conclusions.

\section{Ermakov-Lewis invariant}
In the sixties, Lewis \cite{Lewis0,Lewis} introduced an invariant quantity that has the form (we have set $\hbar=1$ and in this Section dot means derivative with respect to time)
\begin{equation}
 {I}=\frac{1}{2}\left[\frac{ {x}^2}{\rho^2}+\left(\rho  {p}-\dot{\rho} {x}\right)^2\right], \label{1Dinvar}
\end{equation}
with  $\rho$ an auxiliary function that obeys the Ermakov equation \cite{Ray,Thylwe,JPA,Moya2007,Pedrosa}
\begin{eqnarray}\label{Erma}
\ddot{\rho}+\Omega^2(t)\rho=\frac{1}{\rho^3};
\end{eqnarray}
consequently (\ref{1Dinvar}) takes the name Ermakov-Lewis invariant. The operator given in \eqref{1Dinvar} is invariant in the sense that
\begin{eqnarray}
\frac{dI}{dt}=\frac{\partial {I}}{\partial t}+i[  {H},I]=0. 
\end{eqnarray}
It is interesting that even though the time-dependent frequency is zero, i.e. a free particle, the Lewis-Ermakov invariant is maintained for such system. In this case the Ermakov equation reduces to
\begin{eqnarray}\label{ErmaF0}
\ddot{\rho}=\frac{1}{\rho^3},
\end{eqnarray}
and the Hamiltonian is that of the free particle (with mass equal to one), i.e. $H_\text{free}=p^2/2$. Therefore $H_\text{free}$ is not the only invariant for a particle freely evolving but also the explicitly time dependent invariant (\ref{1Dinvar}). From the Lewis-Riesenfeld approach \cite{Riesen}, it follows that the quantum invariant $I$ solves an eigenvalue equation with time-independent eigenvalues and  nonstationary eigenfunctions. Such eigenfunctions are not the solutions of the Schr\"odinger equation, however they may be used to construct such solutions \cite{Riesen}.

For the free particle, the auxiliary Ermakov function takes the form 
\begin{equation}\label{ermasol}
    \rho(t)=(at^2+bt+c)^{\frac{1}{2}},
\end{equation}
that is a solution of equation (\ref{ErmaF0}); \textcolor{black}{with $a,\,b,\,c$ are arbitrary real numbers,} with the constraint $ac-b^2/4=1$. It is worth to mention that for the singular oscillator with constant frequency, there also exists a generalization of the Ermakov invariant which is also explicitly time dependent \cite{ZelayaI,ZelayaII,ZelayaIII}.

\section{Madelung-Bohm approach to paraxial wave propagation}
The optical paraxial propagation equation in one dimension is given by \cite{Marte}
\begin{equation}\label{SE}
i\frac{\partial E(x,z)}{\partial z}= -\frac{1}{2k}\frac{\partial^2 E(x,z)}{\partial x^2}-\frac{1}{2} n^2(x,z)E(x,z),
\end{equation}
where $E(x,z)$ is the propagated field, $k$ is the wave number and $n(x,z)$ is the index of refraction. This equation is formally equivalent to the Schrödinger equation by doing $z\rightarrow t$, $k\rightarrow m$ {\color{black} and $-\frac{1}{2} n^2 \rightarrow V(x,z)$, with $m$ the particle mass, $V(x,z)$ {\it quantum} potential and $n$ the refractive index}. We may paraphrase Feynman \cite{Feynman}, stating that there is always the hope that this point of view will inspire more optical simulations of quantum systems \cite{PRL,Optica,Kapitza}. We may give a solution in terms of a polar decomposition \cite{Hojman,Wyatt,Holland}
\begin{equation}\label{psi}
E(x,z)=A(x,z)e^{i S(x,z)},
\end{equation}
with $A(x,z)$ and $S(x,z)$ real functions that depend on the propagation distance and the position. We may separate the real and imaginary parts that come from the substitution of (\ref{psi}) in (\ref{SE}); the first equation reading \cite{ahms} 
\begin{equation} \label{H-J}
\frac{1}{2k}S'^2-\frac{1}{2}({n_B}^2+n^2)+\dot{S}=0,
\end{equation}
and the second one, the continuity (probability conservation) equation,
\begin{equation}\label{Prob} 
\frac{1}{2k}(2A'S'+AS'')+\dot{A}=0,
\end{equation}
with the Bohm index of refraction defined by \cite{Madelung,Bohm}
\begin{equation}\label{bohmpot}
-\frac{1}{2}{n_B}^2=-\frac{1}{2k}\frac{A''}{A},
\end{equation}
where the dot now represents the  propagation distance derivative and the prime the space derivatives.

\section{Operator  solution of the continuity equation}
It is possible to rewrite Eq.~\eqref{Prob} as a Schrodinger-like equation; for this, we do
\begin{equation}\label{Operator}
\frac{\partial A}{\partial z}=-\frac{1}{2k}\left(2S'\frac{\partial }{\partial x}+S''\right)A,
\end{equation}
that, by using the  operator $\hat{p}=-i\frac{\partial }{\partial x}$,  may be taken to the form
\begin{equation}\label{OperatorI}
\frac{\partial A}{\partial z}=-\frac{1}{2k}\left(i2S'\hat{p}+S''\right)A.
\end{equation}
By choosing 
\begin{equation}\label{ese}
    S(x,z)=Q(x)\dot{\nu}(z)+\mu(z),
\end{equation} 
\textcolor{black}{where $Q, \, \nu$ and $\mu$ are arbitrary well behaved real functions,} such that $S'=Q'\dot{\nu}$, and using the property $[f(x),\hat{p}]=if'(x)$ to rearrange terms, we have
\begin{equation}
2S'\hat{p}=\dot{\nu}(Q'\hat{p}+Q'\hat{p})=\dot{\nu}[Q'\hat{p}+\hat{p}Q'+iQ''];
\end{equation}
so that, we may write Eq. (\ref{OperatorI}) as
\begin{equation}
\frac{\partial A}{\partial z}=-i\frac{\dot{\nu}}{2k}\left(Q'\hat{p}+\hat{p}Q'\right)A.
\end{equation}
The function $\mu(z)$ in equation (\ref{ese}) will define the Gouy phase.
The above equation is readily solvable, with  solution
\begin{equation}
A(x,z)=\exp\left\lbrace -\frac{i}{2k}\int\dot{\nu}(z)dz 
\left[Q'(x)\hat{p}+\hat{p}Q'(x)\right] \right\rbrace A_0(x),
\end{equation}
where $A_0(x)=A(x,t=0)$, the initial condition, is an arbitrary (square integrable) function of position.

\subsection{Propagation of a Gaussian field}
Next, we  assume $Q(x)=x^2/2$ and $k=1$ to find the solution
\begin{equation}\label{0110}
A(x,z)=\exp\left[-i\frac{\nu(z)}{2}\left( x\hat{p}+\hat{p}x\right) \right]A_0(x).
\end{equation}
In the above equation, the operator $\exp\left[ -i\frac{\nu(z)}{2}\left( x\hat{p}+\hat{p}x\right)\right] $
is the so-called squeeze operator \cite{Yuen,Caves,Barnett}. By choosing at $z=0$ the amplitude $A_0(x)=\pi^{-1/4} \exp\left(-x^2/2 \right)$, we obtain after application of the squeeze operator
\begin{equation}
A(x,z)=\frac{1}{\pi^{1/4}}\exp\left[-\frac{x^2}{2}e^{-2\nu(z)}-\frac{\nu(z)}{2}\right],
\end{equation}
where we have used that 
\begin{eqnarray}
    e^{-i\frac{\nu(z)}{2}\left( x\hat{p}+\hat{p}x\right) }xe^{i\frac{\nu(z)}{2}\left( x\hat{p}+\hat{p}x\right) }=x\exp[-\nu(z)],
\end{eqnarray}
that may be easily found from the Hadamard formula that states that for two operators $\hat{A}$ and $\hat{B}$, $e^{r\hat{A}}\hat{B}e^{-r\hat{A}}=\hat{B}+r[\hat{A},\hat{B}]+\frac{r^2}{2!}[\hat{A},[\hat{A},\hat{B}]]+\dots$.\\
Now, we are ready to calculate the Bohm index of refraction, for which we need $A'(x,z)=-xe^{-2\nu(z)}A(x,z)$, so that 
\begin{equation}
A''(x,z)=\left[{x^2e^{-4\nu(z)}}-{e^{-2\nu(z)}}\right]A(x,z),
\end{equation}
to obtain
\begin{equation}\label{0130}
-\frac{1}{2}{n_B}^2(x,z)=-\frac{x^2}{2}\exp\left[-4\nu(z)\right]+\frac{1}{2}\exp\left[-2\nu(z)\right] .
\end{equation}
Equation (\ref{Prob}) implies that $\dot{\mu}=-\exp\left[-2\nu(z)\right] /2$, such that we obtain from  equation (\ref{H-J})
\begin{equation}\label{0140}
-\frac{1}{2}n^2(x,z)+\frac{x^2}{2}\left(\ddot{\nu}+\dot{\nu}^2-e^{-4\nu}\right)=0.
\end{equation}
We change variables as follows: $\rho=\exp\left(\nu\right) $, thus $\dot{\mu}=-1/\left(2\rho^2\right)$, $\dot{\nu}=\dot{\rho}/\rho$ and $\ddot{\nu}=\left( \rho\ddot{\rho}-\dot{\rho}^2\right)/2 $, that produces
\begin{equation}\label{0150}
-\frac{1}{2}n^2(x,z)+\frac{x^2}{2\rho}\left(\ddot{\rho}-\frac{1}{\rho^3}\right)=0.
\end{equation}
The paraxial equation for a freely propagating field is obtained from equation (\ref{ErmaF0}). Its solution, equation (\ref{ermasol}), produces the values
\begin{equation}
 \nu=\frac{1}{2}\ln (az^2+bz +c),\qquad   \mu=-\frac{1}{2}\int \frac{dz}{az^2+bz+c},
\end{equation}
and the conditions $\nu(0)=0 \rightarrow c=1$ and  $a-b^2/2=1$. \textcolor{black}{The choice} $S(x,0)=0=Q(x)\dot{\nu}(0)+\mu(0)$ implies $\nu(0)=\mu(0)=0$, that delivers the values for $a$ and $b$ given by
\begin{equation}
    \dot{\nu}=\frac{1}{2}\frac{2az+b}{az^2+bz +c}, \qquad  \dot{\nu}(0)=0\rightarrow b=0 \, , a=1.
\end{equation}
Therefore, we may write finally the amplitude
\begin{equation}
A(x,z)=\frac{1}{\pi^{1/4}(z^2 +1)^{1/4}}\exp\left[-\frac{x^2}{2(z^2 +1)}\right],
\end{equation}
which is nothing but the amplitude for a Gaussian beam freely (paraxially) propagating. The phase is written as
\begin{equation}
    S(x,z)=\frac{z}{z^2+1}x^2-\frac{1}{2}\int \frac{dz}{z^2+1},
\end{equation}
that gives the Gouy phase
\begin{equation}
    \phi_G(z)=-\frac{1}{2}\arctan z.
\label{gouyphaseknown}
\end{equation}
In this way, we have obtained Gouy's phase from the Madelung-Bohm index of refraction, equation (\ref{0130}), where the quadratic term generates an effective GRIN medium \cite{ChileI} that in turn produces the focusing effect needed to induce the Gouy phase shift. 

\section{Conclusions}
We have used Ermakov-Lewis techniques to show that if we choose a (Bohm) phase that produces a Gaussian field, the Gouy phase arises naturally in the Madelung-Bohm approach to paraxial wave propagation. This is done with the help of the solution of the Ermakov equation coming from the explicitly time dependent invariant introduced by Lewis \cite{Lewis0,Lewis}, in our case for the Hamiltonian of the free particle (paraxial free propagation in classical optics).   \textcolor{black}{Although the Gouy phase \eqref{gouyphaseknown} has been  known for long time \cite{Gouy1,Gouy2} its physical origin is still a matter of research \cite{Michigan,Carmelo}. The procedure detailed here gives a natural explanation of its origin in terms of the form of the spread of the wave, which is measured by the Bohm index of refraction. It is expected that this gives hints on Gouy phases for different kinds of beams.}

\section{Conflict of interest}
The authors declare that they have no conflict of interest.

\section{Author contributions}
The authors declare to have contributed in the same way in the elaboration of this article.

\end{document}